\documentclass{iau}
\usepackage{graphicx}
\newcommand{\teff}{T_{\rm eff}}
\newcommand{\logg}{\log g}

\title[Ensemble Studies of WDs] 
{Ensemble Evolutionary Studies of White Dwarfs in Open Star Clusters}

\author[K.~A.\  Williams]   
{Kurtis A. Williams}

\affiliation{Department of Physics \& Astronomy, Texas A\&M University-Commerce \\ P.O.\ Box 3011, Commerce, TX, 75402, USA \\ email: {\tt Kurtis.Williams@tamuc.edu} }

\pubyear{2020}
\volume{357}  
\date{7 Jan 2020}
\jname{White Dwarfs as probes of fundamental physics and tracers of planetary, stellar, \& galactic evolution}
\editors{M. Barstow, J.L. Provencal, et al., eds.}
\begin{document}

\maketitle

\begin{abstract}
White dwarfs (WDs) in open star clusters are a highly useful ensemble of stars.  While numerous researchers use open cluster WDs to study the initial-final mass relation, numerous other evolutionary studies are also enabled by this sample of stars, including searches for stochastic mass loss, studies of binary star evolution, and measurements of metallicity impacts on WD formation and evolution.  However, it is crucial to use astrometric data such as proper motions to remove contaminating field WDs from open cluster samples; multi-epoch ground based imaging is needed for most open cluster WDs.  Also, the strongly correlated errors in the initial mass - final mass plane must be considered; we illustrate the importance of this consideration using a large open cluster WD sample and Monte Carlo techniques.  
\end{abstract}

\firstsection 
\section{The Utility of Open Cluster White Dwarfs}
Since the start of the millennium, my collaborators and I have 
investigated the white dwarf (WD) populations of open star clusters.  The primary advantage of open cluster WDs is that crucial stellar parameters such as distance, progenitor metallicity, and the total age (WD cooling age plus the nuclear lifetime of the progenitor) are well constrained from observations of non-WD cluster members.  

This additional knowledge opens up avenues of research that are difficult, though not impossible, to pursue with field WDs. Up to now the primary use of open cluster WD samples has been determination of the semi-empirical initial-final mass relation (IFMR).  Early observational studies were headed by Volker Weidemann (e.g., \cite[Weidemann 1977]{Weidemann1977}) and were furthered significantly by Detlev Koester and Dieter Reimers in the 1990s.  Subsequently, the advent of blue-sensitive spectrographs on 8-meter and larger telescopes has allowed several groups to increase the number of known open cluster WDs by a factor of 7 from that in Weidemann's initial study; a recent compilation can be found in \cite{Cummings2018}. This number increases if WDs in field binary systems, which are essentially two-star clusters, are included (e.g., \cite[Catal\'an \etal\  2008]{Catalan2008}).

Over the course of our work (e.g., \cite[Williams \etal\  2009]{Williams2009} and \cite[Williams \etal\  2018]{Williams2018}), my collaborators and I recognized the ever-growing cluster WD sample can shed light on additional stellar evolutionary questions.  For example, the Vogt-Russell theorem is the oft-taught conjecture that the evolutionary path of a single star from a single-metallicity population is determined almost exclusively by its mass, and it is often implicitly invoked in interpretation of WD research.  Yet within NGC 6791 there is significant intrinsic scatter in mass loss (\cite[Kalirai \etal\  2007]{Kalirai2007}), while in M67 no significant scatter in individual WD masses is observed (\cite[Williams \etal\  2018]{Williams2018}).

Multiple star evolution should also affect the WD populations in open star clusters, where binary fractions are high. For example, in M67, we have identified likely WD remnants of blue stragglers, helium-core WD remnants of binary evolution, and a cataclysmic variable (\cite[Williams \etal\  2018]{Williams2018} and \cite[Williams \etal\  2013]{Williams2013}).  Follow-up studies of these WDs should be able to constrain model details such as interaction timescales and remnant system structure.   Such work requires large sample sizes with precision measurements.

Among field WDs, atmospheric layer masses appear to vary from moderately thin (hydrogen layer masses of $\approx 10^{-7} M_\odot$) to the thickest layers possible without igniting a final thermal pulse ($\approx 10^{-4} M_\odot$).  Are these variations a result of stochasticity during the final phases of post main sequence mass loss, or are they due to significant differences in the progenitor stars themselves, such as metallicity? These questions can be addressed via asteroseismic study of open cluster WDs, an idea I first heard from Antonio Kanaan and only recently feasible on large telescopes. 

However, there are many important assumptions and systematic errors plaguing the above studies.  Many of these were quantified by \cite{Salaris2009}.  Below I outline two other critical considerations that must be addressed in future open cluster white dwarf studies: cluster membership and correlated errors in the IFMR.

\section{Cluster Membership Determination}

\begin{figure}[tb]
\begin{center}
 \includegraphics[width=0.6\textwidth]{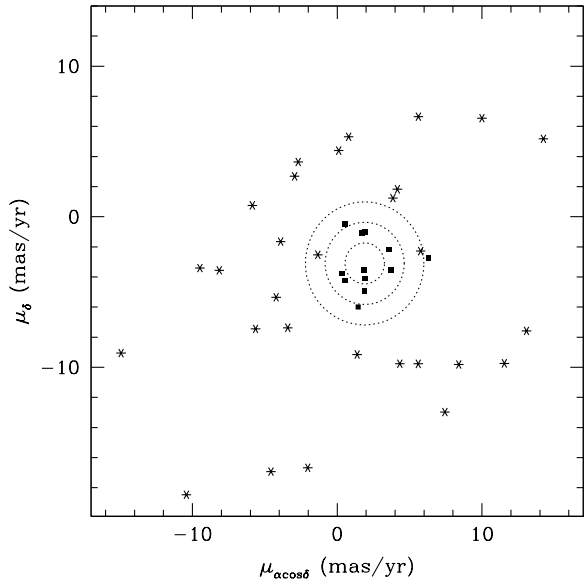} 
 \caption{A simulated proper motion vector diagram for WD candidates in M35, assuming an 11 year baseline and including observational errors. Filled squares are cluster member WDs, and asterisks indicate the field stars.  Dashed circles indicate membership selection criteria of 1.4 mas/yr, 2.8 mas/yr, and 4.2 mas/yr. Even though the cluster WD (filled squares) and field WD (asterisks) proper motions have similar centroids, we can obtain a low-contamination WD sample with our data.}
   \label{fig1}
\end{center}
\end{figure}

Open cluster WD samples require that bona fide cluster member WDs be separated reliably from the significant field WD contamination.  To date the primary means of this us through the use of distance modulus determinations.  The observationally derived $\teff$  and $\logg$ are used to search published WD evolutionary models to calculate the absolute magnitude for each star.  We then calculate the apparent distance modulus for each WD and compare it with the open cluster's apparent distance modulus.  If the two are consistent, and if the cooling age of a WD is younger than the star cluster age, we accept the WD as a likely cluster member.

This method works reasonably well, but it is not a guarantee of cluster membership.  Besancon models (\cite[Robin \etal\ 2003]{Robin2003}) of typical open cluster fields show $\sim 3$ field WDs along the line of sight would meet these membership criteria for cluster ages $\leq 1$ Gyr, a significant contamination when compared to the dozen or fewer cluster member WDs typically identified. Proper motion memberships of WDs in Messier 67 confirms these contamination rates (\cite[Williams \etal\ 2018]{Williams2018}).   Further, distance modulus selection eliminates unresolved double degenerates from cluster WD samples, as they generally appear significantly overluminous.  While this exclusion is desirable for IFMR studies, which assume single-star evolution, it inhibits other areas of open cluster WD research. 

The ideal means to determine cluster membership would be to have parallax and proper motion measurements for each WD.  However, even in the final Gaia catalogs relatively few open cluster WDs will have precision kinematic information due to their intrinsic faintness  ($V\geq 21$ beyond 1 kpc).  We are just now entering the an era where sufficient time has passed since the first epoch of deep, wide-field CCD imaging of open clusters that proper motion memberships for these fainter WDs can be measured by multi-epoch ground-based imaging.

Yet proper motions alone cannot cleanly separate field and cluster WDs.   The young- and intermediate-age clusters targeted in most IFMR studies are dynamically young and co-rotate with the disk field stars, so field and cluster star proper motion distributions are overlapping.  However, the \textit{dispersions} of these distributions are markedly different; field star proper motion dispersions toward kinematically young clusters are $\approx$ $10-15$ mas / yr (\cite[Dias \etal\ 2001]{Dias2001}).  

In order to determine the minimal proper motion precision required to create a clean cluster WD sample, I have simulated ground-based WD proper motion measurements in the field of Messier 35, including the 12 known cluster WDs from \cite[Williams \etal\ (2009)]{Williams2009} and the 29 field WDs identified in that same work.  I randomly drew cluster and field WD proper motions from the values and dispersions in  \cite[Dias \etal\ (2001)]{Dias2001} and included astrometric positional uncertainties of 10 mas per epoch. 

Figure \ref{fig1} shows a representative proper motion plot from this simulation.  Based on 10,000 realizations, we find that a 3 mas/yr selection criteria recovers an average of $\approx 10$ of the 12 cluster member WDs, with an average contamination of $\approx 2$ field WDs.  But this is from proper motion selection \textit{alone}.  Once the distance modulus and age criteria discussed above are applied, the field contamination is less than 1\%. 

\section{Errors in the Semi-Empirical Initial-Final Mass Relation}

\begin{figure}[tbh]
\begin{minipage}{0.48\textwidth}
\centering
\includegraphics[scale=0.38]{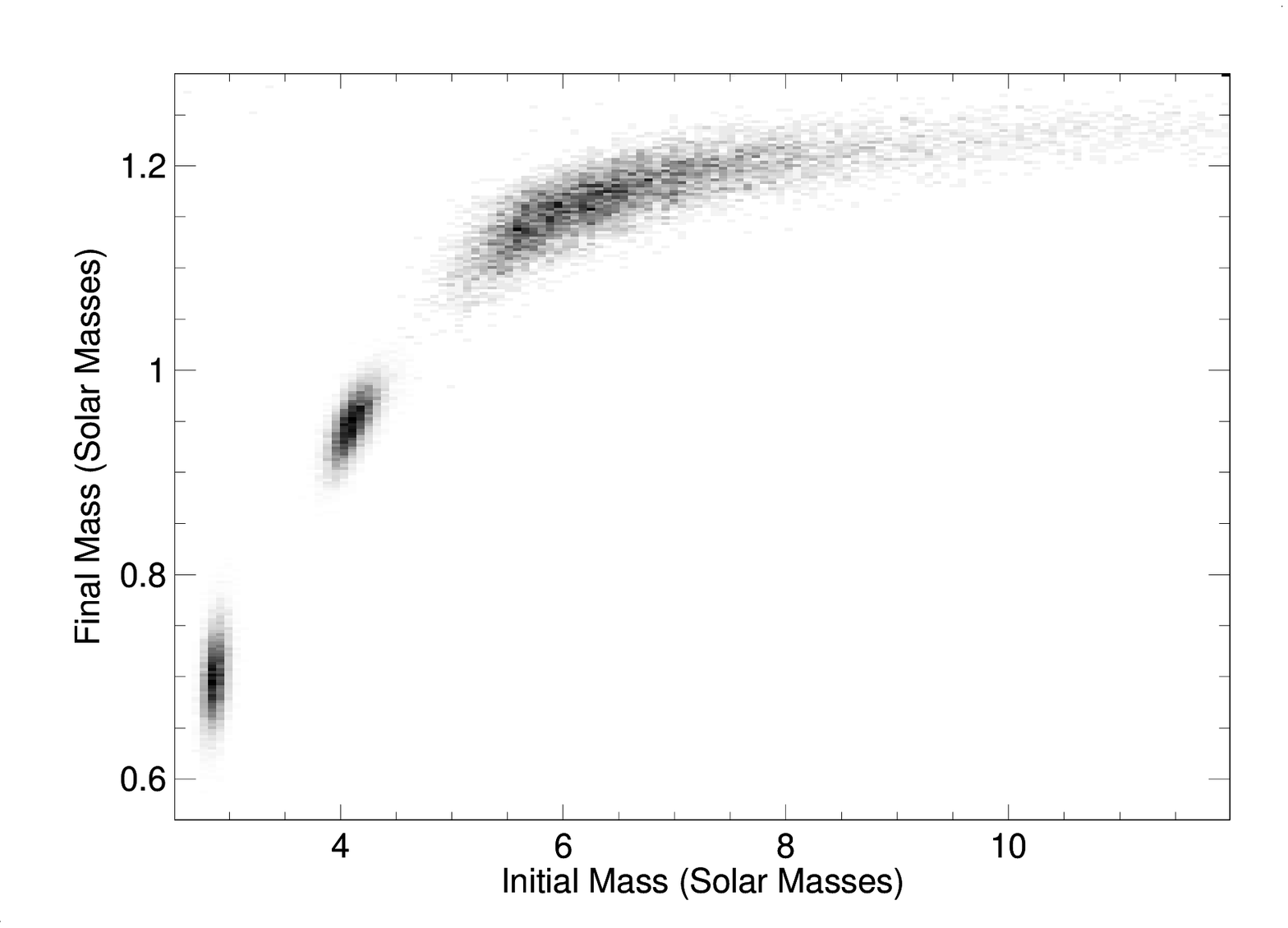}
\end{minipage}
\hfill
\begin{minipage}{0.48\textwidth}
\centering
\includegraphics[scale=0.38]{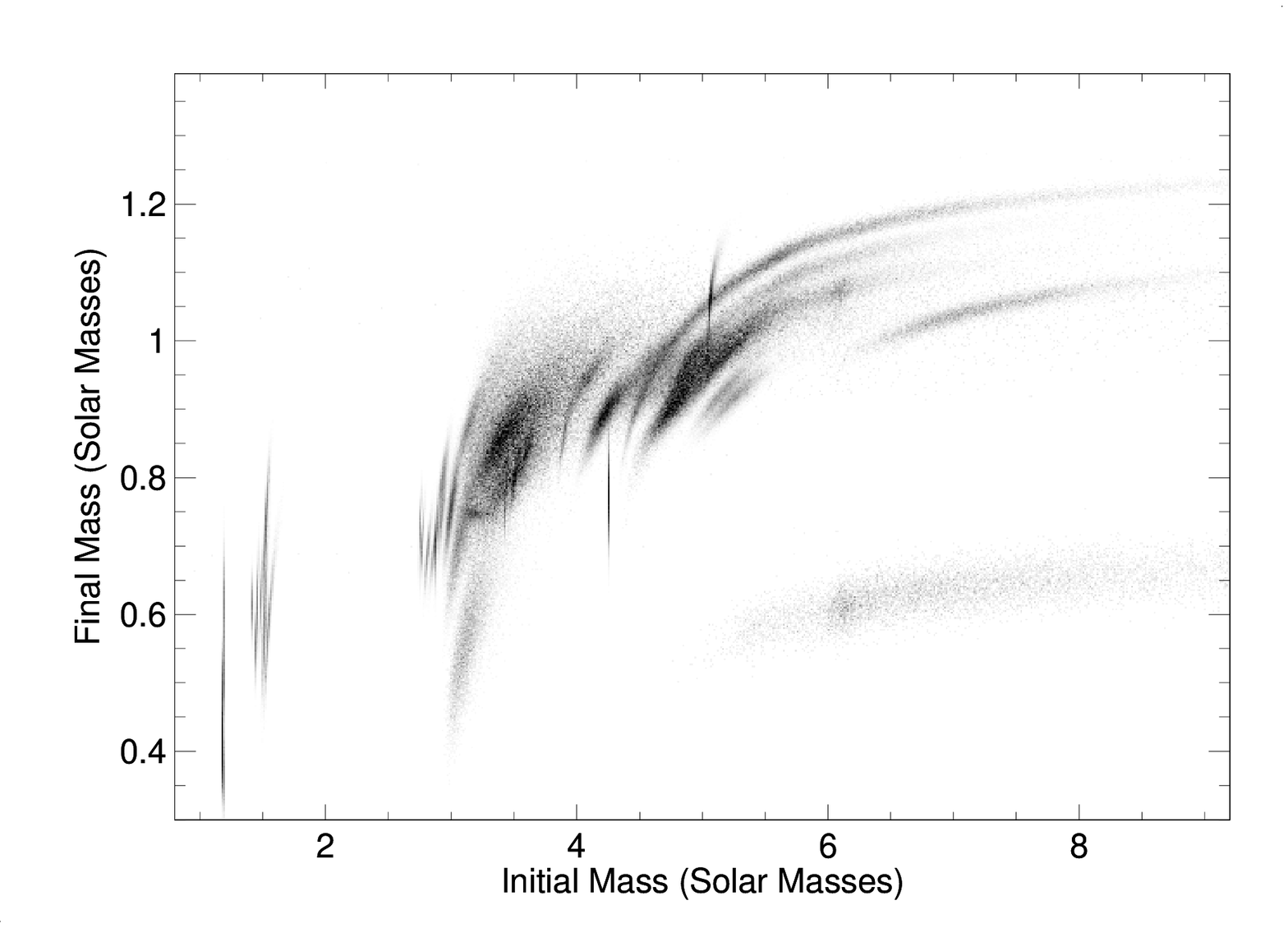}
\end{minipage}
\caption{\textit{Left:} Monte Carlo simulations of the IFMR for three WDs with very high S/N observations: WD 0421+162 (bottom left), NGC 3532 WD 1 (center), and M35 WD 2 (top right). \textit{Right:} Monte Carlo simulations of the current open cluster semi-empirical IFMR from \cite{CantonPhD}. Swooshes indicate the uncertainty regions for individual white dwarfs. All cluster ages are fixed at the values used by \cite[Cummings \etal\ (2018)]{Cummings2018}. Note that orthogonal error bars oft plotted in IFMR determinations are often not representative of the actual uncertainties. \label{fig2}}
\end{figure}

The canonical way of presenting IFMR data and uncertainties is to plot each point with orthogonal error bars in the initial-final mass plane, such as we do in \cite[Williams \etal\ (2018)]{Williams2018}.  However, this practice fails to indicate the underlying highly correlated errors present in the initial-final mass plane, as illustrated in Figure \ref{fig2}.  
The correlated errors are not difficult to understand qualitatively but require numerical methods to quantify.   The observables leading to the IFMR are the $\teff$ and $\logg$ measured from WD spectra.  These measurements are only weakly correlated for the WDs in the current open cluster IFMR and can be ignored, albeit grudgingly.  

The final mass (WD mass) is primarily determined by an assumed mass-radius relationship and the measured $\logg$, with a weak dependence on $\teff$, especially among hotter WDs.   The initial mass (progenitor star mass) is derived by subtracting the WD cooling age from the cluster age; this difference is the nuclear lifetime of the progenitor star.  One can then use a favorite stellar evolutionary model to derive the star's zero-age main sequence mass.  Note the WD cooling age depends strongly on both $\teff$ and $\logg$. 

The quantitative correlation errors in the initial-final mass plane depend on the relative size of the uncertainty in the WD cooling age to the star cluster age and the slope of the progenitor's nuclear lifetime as a function of initial mass.  The left panel of Figure \ref{fig2} shows a Monte Carlo simulation of the initial-final mass relation for three individual WDs with similar $\teff$ and similar observational uncertainties. The shape of the uncertainty distribution is clearly not described by orthogonal error bars in two of the three cases.

In the right panel of Figure \ref{fig2}, I show the uncertainty distributions in the IFMR plane for all of the open cluster WD spectra reanalyzed by Paul Canton in his Ph.D.\ thesis (\cite[Canton 2018]{CantonPhD}).  This sample has significant overlap with that of \cite{Cummings2018} but was obtained independently; the observational parameters measured from the WD spectra are fully consistent within stated errors. This plot strongly suggests that IFMR studies must not ignore the correlated errors of individual points. 

An additional important source of uncertainty in the semi-empirical IFMR is not included in my simulations above but also cannot be ignored, and this is the potential for significant changes in the functional form of the IFMR due to uncertainties in individual star cluster ages.  An error in the assumed cluster age will systematically move all of the WDs in that cluster parallel to the initial mass axis as well as potentially change the shape of each WD's individual uncertainty distribution.  

Most semi-empirical IFMR studies publish the uncertainty in initial masses due to cluster age uncertainties.  However, it is crucial to remember that the changes in initial mass due to these will be systematic, not random errors, within a given cluster. Advanced techniques such as the Bayesian hierarchical modelling of \cite[Si \etal\ (2018)]{Si2018} should therefore be used in IFMR analysis.





\end{document}